\documentclass[10pt]{article}
\usepackage{epsfig}
\begin{document}

\title{\LARGE \bf Superconducting and Spinning Non-Abelian \\
Flux Tubes}
\author{Y. J. Brihaye\thanks{brihaye@umh.ac.be} \\
{ \em Physique Th\'eorique et Math\'ematiques,  Universit\'e de Mons-Hainaut,} \\
{ \em Place du Parc, B-7000  Mons, Belgique}\\
 \setcounter{footnote}{3}
and\\
Y. Verbin\thanks{verbin@openu.ac.il} \\
{ \em Department of Natural Sciences, The Open University
   of Israel,} \\
{ \em P.O.B. 808, Raanana 43107, Israel}}
\date{\ }

\maketitle
\thispagestyle{empty}
\begin{abstract}
 We find new non-Abelian flux tube solutions in a model of $N_f$ scalar fields in the fundamental
 representation of $SU(N)\times U(1)$  with $N \leq N_f$ (the ``extended non-Abelian Higgs model''), 
 and study their main properties. Among the solutions there are spinning strings as well as 
 superconducting ones. The solutions exist only in a non trivial domain of the parameter space defined by the ratio
between the $SU(N)$ and $U(1)$ coupling constants, the scalar self-interaction coupling constants,
the magnetic fluxes (Abelian as well as non-Abelian) and the ``twist parameter'' which is 
a non-trivial relative phase of the Higgs fields.

\end{abstract}

\maketitle
\medskip \medskip

\section{Introduction}

Non-Abelian stringlike solutions have a long history which starts already in 1973 with the
Nielsen-Olesen seminal paper \cite{NO}. Several general discussions were published
\cite{NO,deVega1976,Hasenfratz1979,SchwarzTyupkin}
and explicit solutions (numerical of course) were obtained
\cite{deVegaSchap1986PRL,deVegaSchap1986PRD,HeoVach1998,Suranyi1999,SchaposnikSuranyi,KneippBrockill,KonishiSpanu}
for an $SU(N)$ gauge theory with Higgs fields in the adjoint representation
which completely break the symmetry. However, the major part of the
activity in the field of cosmic strings \cite{VilShel} was concentrated on their Abelian counterparts. One reason for
this is that these non-Abelian string solutions have their flux directed in a fixed direction in the corresponding algebra
so they are essentially Abelian.

In recent years, new kinds of non-Abelian strings were discovered during attempts to understand the 
phenomenon of confinement in QCD \cite{HananyTong2003,Auzzi2003,ShifmanYung2004,HananyTong2004,MarkovEtAl}, and
their properties were studied extensively \cite{EtoEtAlPRL2006,EtoEtAlJPA2006,EtoEtAlPRD2006}. 
These new solutions appear in models with a global (flavor) $SU(N_f)$ symmetry in addition to the 
$SU(N)$ local (color) symmetry based on scalar fields in the fundamental representation. They allow rotation 
of the non-Abelian flux in the Lie algebra, which makes them genuinely non-Abelian. 
When $N_f > N$ these consist a generalization \cite{IsozumiEtAlPRD2005,ShifmanYung2006} 
of the {\it semilocal} strings introduced originally within the extended Abelian Higgs model \cite{VachAch,AchVach}
which is the Higgs system with a global $SU(2)$ symmetry in 
addition to the local $U(1)$. We therefore term the model discussed here the ``extended non-Abelian Higgs model''.

Most of the studies of these non-Abelian string solutions up to now have been limited to the self-dual 
(BPS) case. However, it is natural to go further and look for more general solutions as has just been done 
very recently \cite{GorskyMikh2007,AuzziEtAl2007}. This is the direction which we will take in this  work, 
namely going beyond the BPS limit and it will be done together with allowing also for the possibility of 
rotation (i.e. spinning solutions) and of currents along the string axis. Spinning and superconducting 
cosmic strings have been found recently \cite{FRV1,FRV2} in the extended Abelian Higgs model which gives rise 
also to the (embedded) Nielsen-Olesen solutions. These new semilocal solutions, known as {\it twisted}, 
occur mainly outside the very peculiar (self-dual) 
limit of the coupling constants where the equations of the theory admit Bogomolnyi conditions. 
Another outstanding feature of twisted semilocal strings is that, when they exist, there is a continuous 
family of them, labelled by the ``twist'', a parameter entering through a space-time dependent relative phase 
of the Higgs field components. Note however the existence of the electro-weak superconducting strings 
\cite{Volkov2006} which exist without a ``twist''.

In the Abelian case, the local string is characterized by a magnetic field concentrated
in a tube along the symmetry axis. Outside the core the magnetic field strength decays exponentially. 
The Higgs field  vanishes on the axis and reaches asymptotically its symmetry-breaking value.

\begin{figure}[!t]
\centering
\leavevmode\epsfxsize=4.0cm
\epsfbox{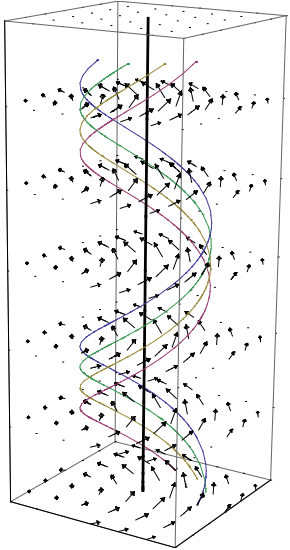}\\
\caption{\label{MagneticField} \small{
The magnetic field and few spiral field lines of a twisted Abelian semilocal string. The thick line is the string axis.}}
\end{figure}

For the twisted semilocal string the geometry is more involved. The  magnetic and Higgs fields behave 
roughly as for local string but the configuration supports in addition an azimuthal (``tangential'') component 
of the magnetic field. The source of this current is the additional Higgs component; 
its modulus is non-zero on the axis (forming a condensate) and vanishes outside the core and its phase is twisted.
The effect of the non trivial phase can be appreciated once computing the gauge invariant Noether currents
and the magnetic field. Figure \ref{MagneticField} gives a pictorial representation of the effect of the twist.

In this work we will present the non-Abelian analogues of the Abelian semilocal twisted strings and study
their properties like current and angular momentum. Untwisted purely magnetic local non-Abelian strings will be 
discussed briefly as a special case. 
The field equations corresponding to the models presented here are non linear and coupled
and do not admit explicit solutions. We therefore rely on numerical techniques to construct the solutions 
and calculate the physical quantities. Several figures are necessary to illustrate the 
extreme richness of the solutions.  

The existence of these new kinds of strings raises the question of the nature of the gravitational 
fields of these strings and the possibility of new features in this respect. Some
initial work has been already done in this area \cite{Aldrovandi2007} -- still for the BPS case only
and we will turn to that question in a future publication. Another issue which is beyond the scope of
this work is the effect of spin on the reconnection probability \cite{HashimotoTong,EtoEtAl2007} of
non-Abelian strings.

This paper has the following plan. In section 2 we present the extended non-Abelian Higgs model and discuss 
its relation with its Abelian counterpart. In section 3 we derive the field equations and obtain the physical 
quantities which are used to characterize the solutions: energy, angular momentum, currents and charges. 
In section 4 we present the various string solutions and discuss their properties across their parameter space. 
Section 5 contains our conclusions.

\section{The extended non-Abelian Higgs model} \label{secExtNAH}
\setcounter{equation}{0}
Our general framework is based on a Lagrangian describing a multiplet of $N_f$ scalar fields
with local invariance under
$SU(N)\times U(1)$ and a global invariance under $SU(N_f)$. The local symmetry further requires $N^2-1$ non-Abelian
gauge fields and one Abelian field.
The scalar fields can be written in term of a matrix with elements
 $\Phi_{as}$ where $1 \leq a \leq N$ and $1 \leq s \leq N_f$ transforming
 according to
\begin{equation}
                 \Phi'_{br} =   U_{ba}(x) \Phi_{as} S^{\dagger}_{sr}
\end{equation}
where $U$ and $S$ are matrices in the fundamental representations 
of $SU(N)$ and $SU(N_f)$ respectively.
The Lagrangian is:
\begin{equation}
\label{lag}
{\cal L} = (D_\mu \Phi_{as})^{*} (D^\mu \Phi_{as})- V(\Phi_{as}) 
-\frac{1}{4} F_{\mu\nu}^a F^{a \mu\nu}
 -\frac{1}{4} F_{\mu\nu} F^{\mu\nu}
\end{equation}
The standard definitions are used  for the covariant derivative
and gauge field strengths~:
\begin{equation}
\label{cova}
D_{\mu} \Phi_s = \Bigl(\partial_{\mu}- ie_1 A_{\mu}
             -ie_2  \tilde{A}_{\mu}^a \tau^a
               \Bigr)\Phi_s
\end{equation}
\begin{equation}
\label{FieldStr}
F_{\mu\nu}=\partial_\mu A_\nu-\partial_\nu A_\mu \;\;\; ; \;\;\;
F_{\mu\nu}^a=\partial_\mu \tilde{A}_\nu^a-\partial_\nu \tilde{A}_\mu^a
            + 2e_2 f^{abc} \tilde{A}_\mu^b \tilde{A}_\nu^c
\end{equation}
where $f^{abc}$ are the structure constants of the gauge group 
and $\frac{1}{2}\tau^a$ are the (hermitian) generators in the
fundamental representation. We use the normalization 
$Tr(\tau^a\tau^b)=2\delta^{ab}$. 

The most general renormalizable symmetry breaking potential is
\begin{equation}
\label{pot1}
     V =\alpha e_1^2(Tr(\Phi^{\dagger}\Phi)-N v^2)^2 + 
     \frac{\beta e_1^2}{2} Tr(\Phi^{\dagger}\tau^a\Phi)Tr(\Phi^{\dagger}\tau^a\Phi)
   \end{equation}
where $\alpha$ , $\beta$  and $v$ are positive constants. 
It can be written also as:
\begin{equation}
\label{pot2}
     V = \alpha e_1^2 (Tr(\Phi^{\dagger}\Phi)-N v^2)^2 + 
     \beta e_1^2 \left( Tr(\Phi^{\dagger}\Phi\:\Phi^{\dagger}\Phi) - 
     \frac{1}{N} ( Tr(\Phi^{\dagger}\Phi))^2 \right)
   \end{equation}
The first term forces the scalar fields to develop
a non-trivial minimum, while the second term forces the
 minimal configurations  $\Phi$ to be such that  $\Phi^{\dagger}\Phi=v^2 I$.

This general system contains some well-known special cases, or seen from the other direction, may be 
considered as a generalization of several systems. From the point of view of the present 
discussion our model generalizes the extended Abelian Higgs model which corresponds to 
$N=1$ (and $e_2 =0$, while $N_f$ is arbitrary) in our terminology. In the Abelian case, the second term 
of the potential vanishes identically. This model allows for two kinds of stable string solutions as 
was first discovered \cite{VachAch} for the case $N_f =2$. The first is just the embedded Abelian 
(Nielsen-Olesen) flux tube which in these circumstances is stable \cite{Hindmarsh1992} for $\alpha\leq 1/2$ .
 The second is sometimes called ``skyrmion'' because of the relation with the $\sigma$-model lumps 
 \cite{Hindmarsh1992} and exists only in the self-dual limit where the masses of the Higgs and the 
gauge particles are equal (or $\alpha = 1/2$). Its stability is guaranteed by a Bogomolnyi-type argument. 
Both solutions are termed semilocal strings and 
their discovery was a surprise at the time, since the vacuum manifold of the model is a $(2N_f-1)$- 
dimensional sphere which does not give rise to non-contractible loops. This discovery gave an 
explicit example for a system where the non-triviality of the first homotopy group of the vacuum 
manifold is not a necessary condition for the existence of stable string-like solutions. In this 
system, the kinetic (gradient) term plays a crucial role since it has a lower symmetry than the 
potential term, and it is this ``mismatch'' between the different symmetries which enables the 
existence of these solutions \cite{Hindmarsh1992,Preskill1992,AchVach}.

The two kinds of solutions mentioned above are static, but stationary spinning semilocal strings also exist 
in this system, as well as solutions which carry a persistent current. These solutions \cite{FRV1,FRV2} belong to 
a new family of solutions named {\it twisted} semilocal strings. The ``twist'' is realized by a relative phase 
between the two Higgs field components, which changes along the string axis and may be also time dependent.
These twisted strings exist for $\alpha > 1/2$ and form a continuous family parametrized by this ``twist'' 
parameter. The main physical property of these twisted strings is the current along the string axis
which exists in both the static and stationary cases and gives rise to azimuthal component of the magnetic field.
Their energy in their rest frame decreases with 
growing current which implies that they are energetically more favorable to be produced during (cosmological) 
phase transitions than the embedded Abelian strings. These new solutions bifurcate with the embedded 
Abelian flux tubes in the limit of vanishing current, thus clarifying the nature of the ``magnetic spreading'' 
instability \cite{Hindmarsh1992,Preskill1992} of the embedded Abelian flux tubes for $\alpha > 1/2$. 

\section{Non-Abelian Semilocal Strings}
\setcounter{equation}{0}

\subsection{Field Equations for Stringlike Solutions} \label{secFEqs}

We look for cylindrically symmetric non-Abelian stringlike solutions and for definiteness, we limit 
ourselves to the case $N_f = N + 1$. But, the ansatz presented below and the corresponding equations can 
be generalized easily. 

The electromagnetic potential and its non-Abelian counterpart will be: 
\begin{equation}
A_\mu dx^\mu = A_0 (r)dt+A_\theta (r)d\theta +A_z (r)dz 
\; ; \;\;\;
 \tau^a \tilde{A}_{\mu}^a dx^\mu 
 = \left( \tilde{A}_0 (r)dt+\tilde{A}_\theta (r)d\theta +
 \tilde{A}_z (r)dz \right)G
 \label{LocalAnsatzMaxYM}
 \end{equation}
where $G$ belongs to the Cartan subalgebra of the gauge group. 
For $SU(2)$ there is only one possibility, i.e. $G=\sigma_3$
but already for $SU(3)$ which is a rank 2 group there is more freedom 
and $G$ is a linear combination of the two (commuting) diagonal Gell-Mann matrices.
We will denote the diagonal elements of $G$ by $G_a$ with $a=1,...,N$.

For the scalar field we take:
\begin{eqnarray}
    \Phi &=& v\left(
      \begin{array}{ccccccccc}
     \varphi_1(r)e^{i\alpha_1} & 0 & \cdots & 0 & 0 & 0\\
    0& \varphi_2(r)e^{i\alpha_2} & \cdots & 0& 0 & 0\\
          \vdots & \vdots & \ddots & \vdots& \vdots &\vdots& \\
          0& 0 & \cdots & 0& \varphi_{N}(r)e^{i\alpha_{N}}, &\rho_N(r)e^{i\chi_{N}} \\
          \end{array}
          \right)
     \label{SemilocalAnsatz}
\end{eqnarray}
which follows (and actually generalizes) the form of \cite{Auzzi2003} or \cite{Aldrovandi2007}.
The phases $ \alpha_i $ and $\chi_{N}$ will have linear dependence on $t$, 
$\theta$ and $z$: $\alpha_i = \omega_i t + m_i \theta +k_i z$ and 
$\chi_{N} = \varpi t + n \theta +q z$. 

 To write the field equations, it is convenient to use
 a dimensionless coordinate, to rescale the fields and ``phase parameters''
 appropriately and to define a relative gauge coupling constant:
\begin{eqnarray}
 \label{scaling}
  r = \frac{x}{v e_1} \ \ , \ \  \  
  (\omega_i,k_i) \to v e_1  (\omega_i,k_i) \ \ , \ \ 
  (\varpi,q) \to v e_1  (\varpi,q)  \ \ , \ \ \delta = \frac{e_2}{e_1} \\ \nonumber  
  ( A_{0}(r),A_{z}(r)) = v ( A_{0}(x),A_{z}(x)) \ \ , \ \
  (\tilde A_{0}(r),\tilde A_{z}(r))= \frac{v e_1}{e_2} (\tilde A_{0}(x),\tilde A_{z}(x))
\end{eqnarray}
With these dimensionless quantities, the field 
equations of the system are written below. For the scalar fields we obtain (recall: $G_a$ are 
the diagonal elements of $G$)~:
\begin{eqnarray}
\frac{(x\varphi'_a)'}{x}=\left( \frac{(m_a - A_\theta - \tilde{A}_\theta G_{a})^2}{x^2}-
(\omega_a - A_0 - \tilde{A}_0 G_{a})^2+
(k_a - A_z - \tilde{A}_z G_{a})^2 \right) \varphi_a  \label{EqscalarSemilocal} \\ \nonumber
+ 2\left[ \alpha \left(\rho_N^2 + \sum_{b} \varphi_b^2 - N \right) +\frac{\beta}{N} 
\left( (N-1)\varphi_a^2 - \sum _{b\neq a} \varphi_b^2 -\rho_N^2 \right) \right] \varphi_a \;\;\;, a<N
\end{eqnarray}

\begin{eqnarray}
\frac{(x\varphi'_N)'}{x}=\left( \frac{(m_N - A_\theta - \tilde{A}_\theta G_{N})^2}{x^2}-
(\omega_N - A_0 - \tilde{A}_0 G_{N})^2+
(k_N - A_z - \tilde{A}_z G_{N})^2 \right) \varphi_N \\ \nonumber
+2 \left[ \alpha \left(\rho_N^2 + \sum_{b} \varphi_b^2 - N \right) +\frac{\beta}{N} 
\left((N-1)(\rho_N^2 +\varphi_N^2)  - \sum _{b=1}^{N-1} \varphi_b^2 \right) \right] \varphi_N 
\label{EqfNSemilocal}
\end{eqnarray}

\begin{eqnarray}
\frac{(x\rho'_N)'}{x}=\left( \frac{(n - A_\theta - \tilde{A}_\theta G_{N})^2}{x^2}-
(\varpi - A_0 - \tilde{A}_0 G_{N})^2+
(q - A_z - \tilde{A}_z G_{N})^2 \right) \rho_N \\ \nonumber
+2 \left[ \alpha \left(\rho_N^2 + \sum_{b} \varphi_b^2 - N \right) +\frac{\beta}{N} 
\left((N-1)(\rho_N^2 +\varphi_N^2)  - \sum _{b=1}^{N-1} \varphi_b^2 \right) \right] \rho_N 
\label{EqRhoSemilocal}
\end{eqnarray}
where there is no summation on repeated indices and all sums are for $a,b = 1,...,N$
unless indicated otherwise. For the Abelian-gauge fields, we have
\begin{eqnarray}
\frac{(xA'_0)'}{x}=-2\sum_{a} (\omega_a - A_0 - \tilde{A}_0 G_{a}) \varphi_a^2 -2(\varpi - A_0 -
\tilde{A}_0 G_{N})\rho_N^2
\label{MaxwellEqSemilocal0}\\
x(A'_\theta /x)'=-2\sum_{a} (m_a - A_\theta - \tilde{A}_\theta G_{a}) \varphi_a^2 -
2(n - A_\theta - \tilde{A}_\theta G_{N})\rho_N^2
\label{MaxwellEqSemilocalTH}\\
\frac{(xA'_z)'}{x}=-2\sum_{a} (k_a - A_z - \tilde{A}_z G_{a}) \varphi_a^2 -2(q - A_z - 
\tilde{A}_z G_{N})\rho_N^2
\label{MaxwellEqSemilocalZ}
\end{eqnarray}
Finally, for the non-Abelian fields~:
\begin{eqnarray}
\frac{(x\tilde{A}'_0)'}{x}=-2\delta^2 \left( \sum_{a} (\omega_a - A_0 - 
\tilde{A}_0 G_{a}) \varphi_a^2 G_{a} + (\varpi - A_0 - \tilde{A}_0 G_{N})\rho_N^2 G_{N} \right) \\
x(\tilde{A}'_\theta /x)'=-2\delta^2 \left( \sum_{a} (m_a - A_\theta - 
\tilde{A}_\theta G_{a}) \varphi_a^2 G_{a} + (n - A_\theta - \tilde{A}_\theta G_{N})\rho_N^2 G_{N} \right)\\
\frac{(x\tilde{A}'_z)'}{x}=-2\delta^2 \left( \sum_{a} (k_a - A_z - \tilde{A}_z G_{a}) \varphi_a^2 G_{a} + 
(q - A_z - \tilde{A}_z G_{N})         \rho_N^2 G_{N} \right)
\label{YMeqsSemilocal}
\end{eqnarray}

In order to obtain localized solutions, appropriate boundary conditions should be imposed. 
The boundary conditions will guarantee that the solutions will be
regular on the axis $x=0$ and approach a vacuum for $x \to \infty$. Regularity at $x=0$ 
yields 
\begin{eqnarray}
\varphi_a (0)=0 \ ,\ A_\theta (0)=0 \ ,\ \tilde{A}_\theta (0)=0 
\ ,\ \rho'_N(0) = 0 \ ,\  A'_{0,z}(0) =0 \ ,\ \tilde A'_{0,z}(0)=0  
\label{BCZero} 
\end{eqnarray}
while the requirement for asymptotic approach to the vacuum is somewhat more
involved. For the scalar field we simply have 
\begin{equation} 
\label{BCScalarInf}
 \varphi_a(\infty) = 1 \ , \ \rho_N(\infty) = 0 \ \
\end{equation}
which makes a clear distinction between the $N$ fields $\varphi_a$ and the 
$\rho_N$ field. This last one will either trivially vanish, or have a unique behavior
of starting at a non-vanishing central value and decreasing monotonically to zero.
Therefore, to avoid singularity at $x=0$ all the solutions discussed 
in the following will have $n=0$.

For the asymptotic behavior of the gauge fields we start by the observation
that under a Lorentz boost in the $z$ direction, the quantities
$(\omega_i,k_i)$ transform as Lorentz two-vectors. Since we will
consider in this paper only the case when these two vectors are
spacelike (corresponding to the magnetic case in the classification
of \cite{FRV2}), we can choose {\it one} of these
two-vectors to be of the form $(0,k_i)$ without loosing generality.
Independently, one can perform  residual gauge transformations
of the form
\begin{equation}
              U_{res} = \exp i( (at+bz)I + (a't+b'z)G)
\end{equation}
and assume  in the following
\begin{equation}
\label{gaugeres}
A_0(\infty)=0\ ,\ A_z(\infty)=0\ ,\ \tilde A_0(\infty)=0\ ,\ \tilde A_z(\infty)=0.
\end{equation}
As for the azimuthal components of the gauge fields we have to impose
\begin{eqnarray}
m_a - A_\theta (\infty) - \tilde{A}_\theta (\infty)G_{a} =0 
\label{BCThetaInf}
\end{eqnarray}
These are $N$ equations for the two unknowns  $A_\theta (\infty)$
and $\tilde{A}_\theta (\infty)$, so the three magnetic numbers $m_a$ cannot be independent. Moreover, 
since they must be integers, not every $G$ in the Cartan subalgebra is possible. 
A closed formula for these relations is rather complicated so we demonstrate this by two examples 
for the case $N = 3$ where we can parametrize $G$ by: $G=\lambda_3 \cos\psi + \lambda_8 \sin\psi$
where $\lambda_i$ are Gell-Mann matrices. For $\psi = \pi/6$ we have the relation $m_2=m_3$ and 
$\tilde{A}_\theta (\infty) = (m_1-m_3)/\sqrt{3}$,
while for $\psi = \pi/3$ we have  $m_2=(m_1+m_3)/2$ and accordingly 
$\tilde{A}_\theta (\infty) = (m_1-m_3)/2$. $A_\theta (\infty)$ however is given always by 
$A_\theta (\infty) = (m_1+m_2+m_3)/3$.
The concrete solutions which we will present will be mainly with $N = 3$ (other values of
$N$ can be treated similarly).

\subsection{Physical quantities: energy, angular momentum, currents and charges}

In this section, we express several quantities which characterize
the different types of solutions. The quantities are presented for string solutions of the
extended non-Abelian Higgs model. The corresponding formulas for the ``non extended'' model
(i.e. $N_f =N$) are obtained by setting $\rho_N =0$ and $\chi_N =0$ in  Eq. (\ref{SemilocalAnsatz})
and those that follow.
 
The energy and angular momentum (per unit length) of the solutions are obtained by computing the integrals 
\begin{equation}
E=2\pi v^2 \int_0^\infty dx x ({\cal E}_0 +{\cal E}_1 + {\cal V}) \;, \;\;\;\ J=2\pi v^2 \int_0^\infty dx x {\cal J}
\label{EnergyAndSpin} 
\end{equation}
The energy densities ${\cal E}_0,{\cal E}_1,{\cal V}$, correspond respectively to the
scalar part, the gauge field part and the potential. They are given by
\begin{eqnarray}
{\cal E}_0 =  
\sum_{a}\left[\varphi_a'^2+\left( \frac{(m_a - A_\theta - \tilde{A}_\theta G_{a})^2}{x^2}+
(\omega_a - A_0 - \tilde{A}_0 G_{a})^2+ (k_a - A_z - \tilde{A}_z G_{a})^2 \right) \varphi_a^2 \right] \\ \nonumber
+\rho_N'^2 + \left( \frac{(n - A_\theta - \tilde{A}_\theta G_{N})^2}{x^2}+
(\varpi - A_0 - \tilde{A}_0 G_{N})^2+ (q - A_z - \tilde{A}_z G_{N})^2 \right)\rho_N^2 \nonumber
\label{EnergyDensity0}
\end{eqnarray}
\begin{eqnarray}
{\cal E}_1 = \frac{1}{2}\left(A_0'^2 + (A'_\theta /x)^2 + A_z'^2\right) + 
\frac{1}{2\delta^2}\left(\tilde{A}_0'^2 + (\tilde{A}'_\theta /x)^2 + \tilde{A}_z'^2\right)
\label{EnergyDensity1} 
\end{eqnarray} 
\begin{eqnarray}
{\cal V} = \alpha \left(\sum_a \varphi_a^2 + \rho_N^2 -N \right)^2 
+ \beta \left[\sum_a \varphi_a^4 + 2 \rho_N^2 \varphi_N^2 + \rho_N^4 - 
\frac{1}{N}\left(\sum_a \varphi_a^2 + \rho_N^2 \right)^2\right]
\end{eqnarray}
The angular momentum density ${\cal J}$ has the form 
\begin{eqnarray}
{\cal J}&=&2\sum_{a}(\omega_a - A_0 - \tilde{A}_0 G_{a})(m_a - A_\theta - \tilde{A}_\theta G_{a})\varphi_a^2 \\ \nonumber
&+&2(\varpi - A_0 - \tilde{A}_0 G_{N})(n - A_\theta - \tilde{A}_\theta G_{N})\rho_N^2 + 
A'_0 A'_\theta + \frac{1}{\delta^2}\tilde{A}'_0 \tilde{A}'_\theta
\label{SpinDensity} 
\end{eqnarray}

The solutions are also characterized by the Noether charges and currents associated with the global $SU(N+1)$ 
symmetry of the Lagrangian.
The charges (per unit length) are the integrals of the time components of the current densities
\begin{equation}
j_{\mu}(R) = -i \left( \Phi_{ap}^{*}R_{pq} D_{\mu} \Phi_{aq} -  (D_\mu \Phi_{ap})^{*} R_{pq} \Phi_{aq} \right)
\label{currents} 
\end{equation}
where $R$ denotes any of the generators of the global symmetry and the summation convention is used for both the local 
and global indices. For the ``diagonal'' ansatz, Eq. (\ref{SemilocalAnsatz}), we get the following expressions
for the global charge densities 
\begin{equation}
j_{0}(R)=2 \sum_{a}R_{a}(\omega_a - A_0 - \tilde{A}_0 G_{a})\varphi_a^2 + 
2R_{N+1}(\varpi - A_0 - \tilde{A}_0 G_{N})\rho_N^2
\label{GlobalChargeDens} 
\end{equation}
where we denote by $R_s$ the diagonal elements of the commuting flavor 
generators as we do for local symmetry. Similarly for the current densities: 
\begin{equation}
j_{z}(R)=2 \sum_{a}R_{a}(k_a - A_z - \tilde{A}_z G_{a})\varphi_a^2 + 
2R_{N+1}(q - A_z - \tilde{A}_z G_{N})\rho_N^2 \ \ .
\label{GlobalCurrentDens} 
\end{equation}
Accordingly, the charges and currents will be
\begin{equation}
Q(R)=2\pi  \int_0^\infty dx x j_{0}(R)=\sum_{s=1}^{N+1}R_{s}Q_{s} \;, \;\;\;\ 
I(R)=2\pi  \int_0^\infty dx x j_{z}(R)=\sum_{s=1}^{N+1}R_{s}I_{s}
\label{ChargesAndCurrents} 
\end{equation}
where the ``single field contributions'' $Q_{s}$ and $I_{s}$ are defined in an obvious way. In
terms of these, the angular momentum $J$ of the string can be written as $J = \sum m_a Q_a+nQ_{N+1}$.

Since the non-Abelian charges all vanish due to the boundary conditions, these ``single field 
contributions'' are not independent but satisfy the following identities among the charges and currents
\begin{equation}
            \sum_{s=1}^{N+1} Q_s = \sum_{s=1}^{N+1} I_s = 0 \ \ , \ \ 
            \sum_{a=1}^{N} G_{a}Q_a + G_{N}Q_{N+1}= \sum_{a=1}^{N} G_{a} I_a + G_{N}I_{N+1}= 0.
\label{SumChargesAndCurrents}             
\end{equation}
These  identities allow one to express two of the currents (or charges) in terms of the others.
As a consequence, for an $SU(N+1)$ global symmetry, the $N$ diagonal global currents can be expressed in terms
of $N-1$ independent ones. For example, for local $SU(2)$ and global $SU(3)$, these relations lead to $I_1=0$; 
 therefore $I_2=-I_3$ by the left equation of (\ref{SumChargesAndCurrents}) and the two diagonal global currents 
 are proportional to $I_3$.

 For local $SU(3)$ and the choice $\psi = \pi/6$, the global $SU(4)$ currents depend on two of the 
``single field'' currents $I_{s}$ say, $I_{2}$ and $I_{4}$. Actually two of them are just proportional to 
$I_{2}$ and $I_{4}$ and the third is a combination:
\begin{equation}
I(R_3)=-2I_2  \;\;\; ; \;\;\; I(R_8)=-\frac{1}{\sqrt{3}}(3I_2 +2I_4 ) \;\;\; ; \;\;\; I(R_{15})=-\sqrt{\frac{8}{3}}I_4.
\label{GlobalCurrents}             
\end{equation}
These identities provide useful crosschecks of the numerical solutions discussed next.

\section{Discussion of the solutions}
\setcounter{equation}{0}

We solved the system of non-linear equations (\ref{EqscalarSemilocal})-(\ref{YMeqsSemilocal})
by using a numerical routine based on the collocation method \cite{COLSYS}.
The solutions were constructed with grids involving typically 1000 points
and with accuracy of the order of $10^{-8}$. The solutions we found are described
in the following.

\subsection{Special Case: Local strings} \label{secLocalStrings}

Before discussing the semilocal strings, we pause to describe the simplest kind of non-Abelian
ones, namely the {\it local} non-Abelian strings. They are obtained by truncating the general 
system to the case with $N_f = N$ so $\rho_N =0$ and $\chi_N =0$. This requires the 
further substitutions $\varpi=n=q=0$ in the field equations above.

\begin{figure}[!t]
\centering
\leavevmode\epsfxsize=10.0cm
\epsfbox{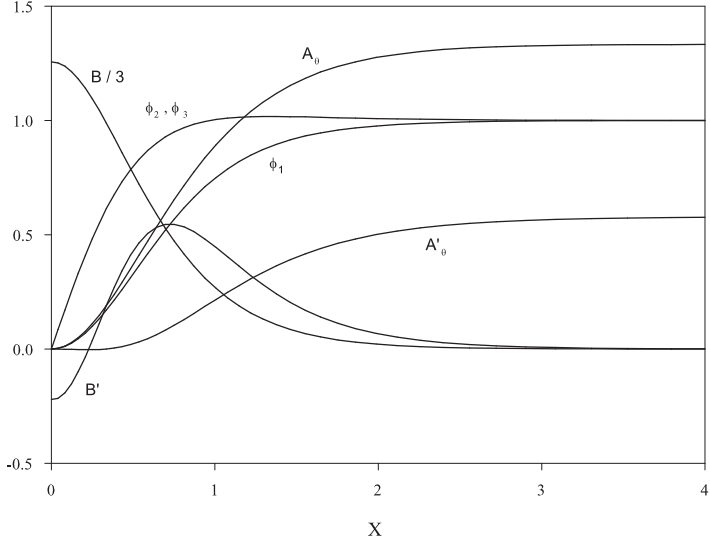}\\
\caption{\label{fig1} \small{
A non-Abelian $SU(3)$ local string: Profiles of the three 
Higgs fields $\varphi_1$, $\varphi_2$, $\varphi_3$, the two
gauge fields $A_\theta$ and $\tilde{A_\theta}$ ($A'_\theta$ in the figure) and the 
corresponding magnetic fields for $m_1=2$,
 $m_2=1$,  $m_3=1$ . This choice yields the relation $\varphi_2=\varphi_3$. 
The Abelian flux is 3/2; the non-Abelian flux is $1/\sqrt{3}$. 
 The relative gauge strength is $\delta=1$ and the parameters in the potential are: $\alpha=\beta=1$.}}
\end{figure}

Inspection of the boundary conditions for the gauge fields as $x\to \infty$, reveals
that in the gauge choice (\ref{gaugeres}) all the parameters $\omega_j,k_i$
should vanish, leading to $A_0(x)=A_z(x)= \tilde A_0(x) = \tilde A_z(x)=0$.
So we solve equations (\ref{EqscalarSemilocal})-(\ref{YMeqsSemilocal}) 
with the above substitutions and with the following boundary conditions : 
\begin{eqnarray}
\varphi_a (0)=0 \;,& A_\theta (0)=0 \;,\;\; \tilde{A}_\theta (0)=0  \label{BCZeroLocal} \\
\varphi_a (\infty)=1 \;,& m_a - A_\theta (\infty) - \tilde{A}_\theta (\infty)G_{a} =0 
\label{BCInf}
\end{eqnarray}
The only stringlike solutions in this system are of the magnetic type, labelled by the integers $m_a$
which are consistent with the value we pick for the angle $\psi$ in the 3-8 plane of the
Cartan subalgebra.

In Fig. \ref{fig1} we show the field profiles of a typical $SU(3)$ local string solutions with 
$\psi = \pi/6$. The Abelian flux is still quantized but is not an integer as can be easily seen from the figure. 
Its value is 3/2. Similarly, the non-Abelian flux is $1/\sqrt{3}$.

\subsection{Semilocal strings}
Now we return to the general case of the full system with $N_f = N+1$, that is equations 
(\ref{EqscalarSemilocal})-(\ref{YMeqsSemilocal}) with non-vanishing 
$\rho_N $ and $\chi_N $. The solutions have in general $(\varpi,q) \neq 0$ 
while we can still exploit the remaining symmetry to require again 
$\omega_j=k_j=0$ for $j=1,\dots,N$. 

In contrast with the local case, the equations for the fields
$A_0$, $A_z$, $\tilde A_0$ and $\tilde A_z$ are not trivially satisfied if $\rho_N(x) \neq 0$
and we have now non-trivial boundary conditions for these gauge components. The boundary conditions are 
contained in equations (\ref{BCZero}), (\ref{BCScalarInf}), 
(\ref{gaugeres}) and (\ref{BCThetaInf}) and we will not repeat them here. We just
point out that the scalar and gauge fields reach their asymptotic
   values with exponential corrections. In particular, the field $\rho_N$ decays according to 
   $\rho_N \propto \exp{[-\sqrt{q^2-\varpi^2}\,\ x]}$, which explains why this field can decay 
   slower than the others  for  $\sqrt{q^2-\varpi^2} << 1$. We have therefore
to solve for  $\rho_N (x)$, $A_{0}(x)$, $A_{z}(x)$, $ \tilde A_{0}(x)$ and $ \tilde A_{z}(x)$ 
in addition to the $N$ scalar components and the two gauge components $A_\theta$ and $\tilde{A_\theta}$.

Along with \cite{FRV1}, we call solutions of this type ``twisted'' solutions. The parameter $q$ 
will be referred to as the ``twist'' parameter. The solutions will be generally ``twisted'', but special ``untwisted''
solutions (with $q=0$) also occur. We also distinguish between static solutions with $\varpi = 0$ and 
stationary for $\varpi \neq 0$. Their properties will be studied below.

\subsection{Purely magnetic (``untwisted'') solutions} \label{pms}
The simplest kind of semilocal solutions is the {\it embedded} (non-Abelian) local flux tubes. These are purely 
magnetic solutions of the full system with $\rho_N =0$. It is therefore evident that
$A_0=A_z=\tilde{A}_0=\tilde{A}_z=0$ and the non-vanishing fields, the $N$ scalars $\varphi_a$,
and the two gauge components $A_\theta$ and $\tilde{A_\theta}$, are the same as for the local strings of 
Sec. \ref{secLocalStrings}.

Another kind of purely magnetic strings is the {\it untwisted} semilocal
strings which generalize the ``Skyrmion'' solutions of the Abelian model mentioned in Sec. 
\ref{secExtNAH}. These solutions are only self-dual as in the Abelian case, and are contained
already among the results of earlier studies \cite{IsozumiEtAlPRD2005,ShifmanYung2006,Aldrovandi2007}. 
In our parametrization they exist for $\alpha=1/2, \beta = 1, \delta = 1$.

The more general kinds of solutions which contain ``twists'' \cite{FRV1,FRV2} 
will be discussed in the next sections.

\subsection{Twisted static and stationary Semilocal strings}

The space of twisted semilocal strings decomposes into three regions which are labelled 
according to the sign of the norm of the two-vector $(\varpi,q)$, that is according to the
sign of $\varpi^2-q^2$. In the ``timelike'' region ($\varpi^2-q^2>0$) we were unable to find localized solutions
as happens also in the Abelian case \cite{FRV2} (recall also the asymptotic behavior mentioned above), 
so we do not discuss them further. 
Solutions with $\varpi^2-q^2=0$ are known as ``chiral'' and will be 
discussed below among the stationary solutions. If we assume the two-vector $(\varpi,q)$ to be spacelike, 
we can set $\varpi=0$ by an appropriate Lorentz boost. The field equations for $A_0$ and $\tilde A_0$ are 
satisfied trivially by $A_0(x)=\tilde A_0(x)=0$ and all the charges $Q_s$ vanish with this choice. The 
components $A_z$ and $\tilde A_z$ as well as the currents $I_s$ do not vanish.

  \begin{figure} [!b]
        \includegraphics[height=.28\textheight,width=.52\textwidth]{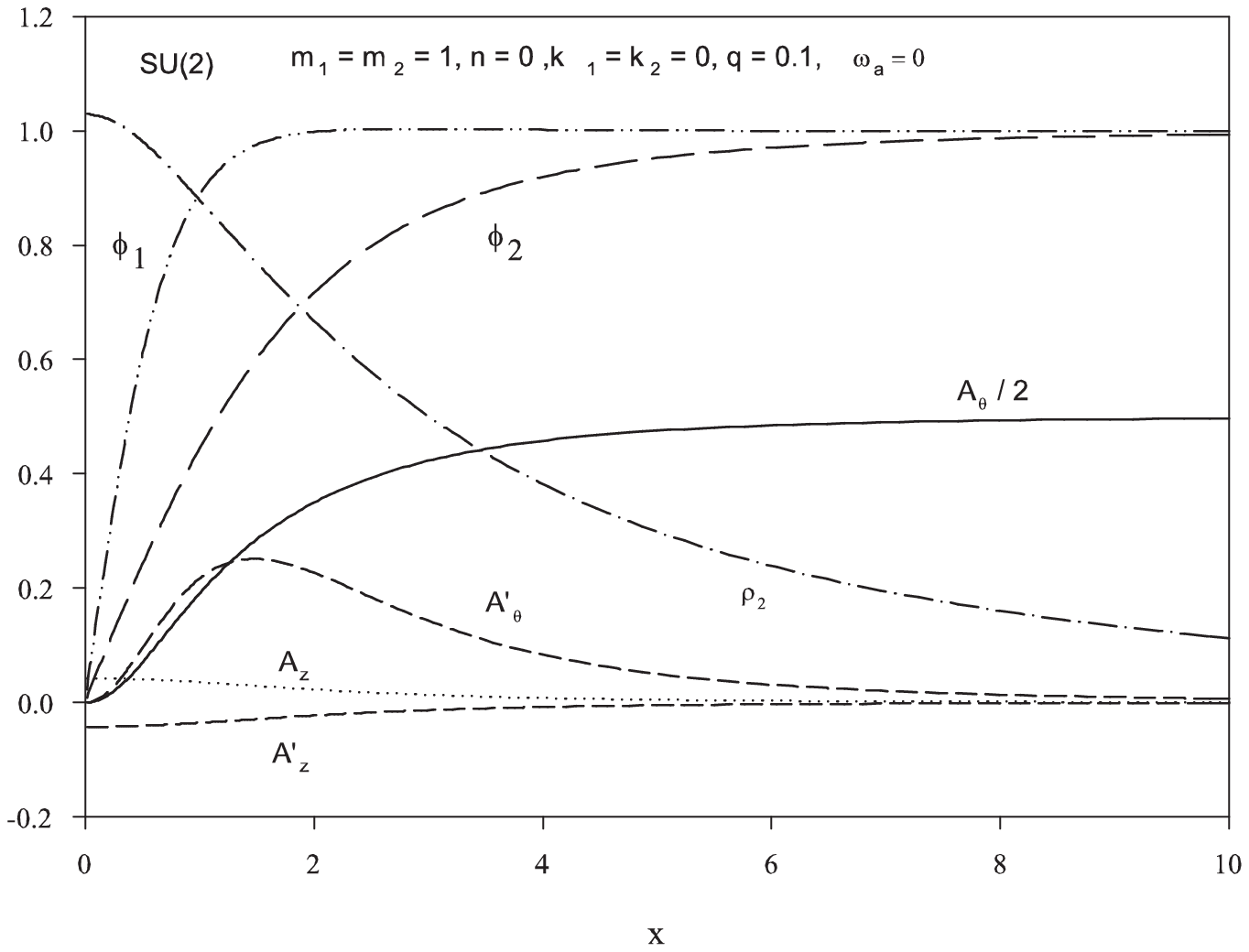}  
      \includegraphics[height=.28\textheight,width=.52\textwidth]{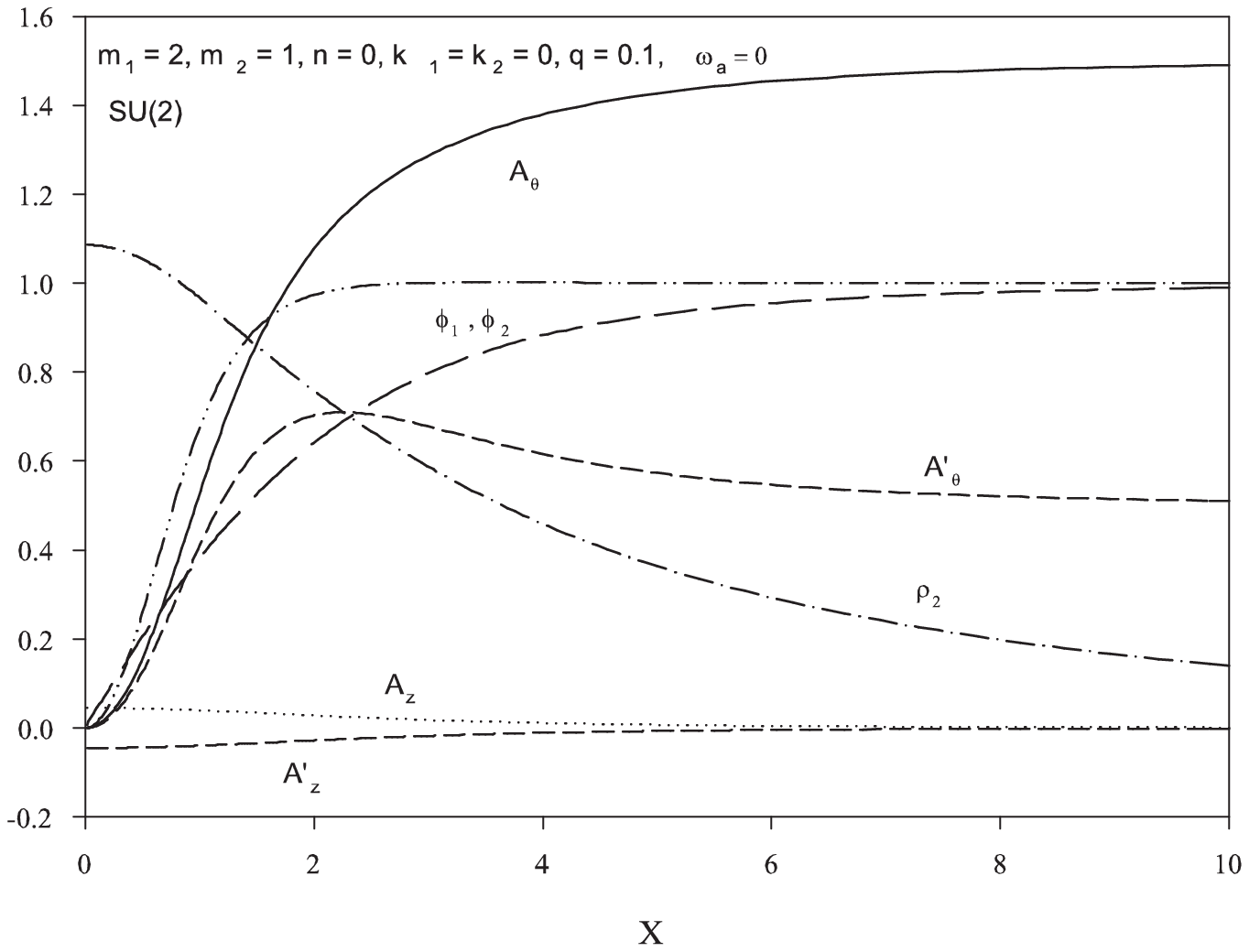}\\
   \caption {\label{figsStaticTWSU2} \small{A non-Abelian $SU(2)$ static twisted 
   semilocal strings: Profiles of the  three Higgs fields $\varphi_1$, $\varphi_2$, $\rho_2$ 
   and the components of the two gauge fields $A_\theta, A_z$ and 
 $\tilde{A_\theta}, \tilde{A_z}$ ($A'_\theta$, $A'_z$ in the figure). 
 The other parameters are: $\alpha=1$, $\beta=\frac{1}{2}$, $\delta=1$.
 Left: Note that the  non-Abelian magnetic flux vanishes and that 
  $A_z = -\tilde{A_z}$. The Abelian flux is 1.
 Right:  Note that  $A_z = -\tilde{A_z}$.  The Abelian flux is 3/2; the non-Abelian flux is $1/2$.}} 
\end{figure} 

\begin{figure}[!t]
\centering
\leavevmode\epsfxsize=10.0cm
\epsfbox{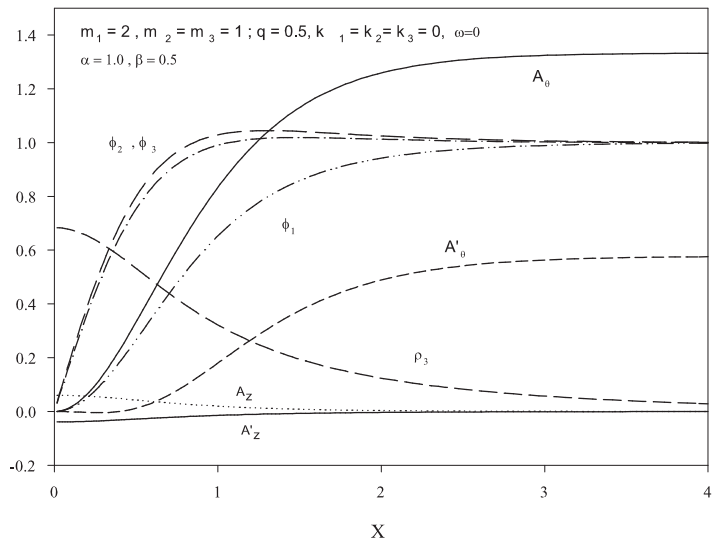}\\
\caption{\label{fig6} \small{A non-Abelian $SU(3)$ static twisted 
   semilocal string ($A'$ stands for $\tilde{A}$). $\delta=1$.}}
\end{figure}

In the Abelian case ($N=1$ and $N_f=2$ with irrelevant $\beta$) \cite{FRV1,FRV2}
twisted semilocal strings appear for large enough values of the potential coupling constant $\alpha$ 
(precisely for $\alpha > 1/2$ in our notations). They form continuous family of static solutions and are 
labelled by the ``twist parameter'' $q$ appearing in the phase of the scalar function $\rho_2$; this parameter takes values
in a finite interval, that is $q \in [0, q_{cr}]$ where $q_{cr}$ is the maximal twist where bifurcation
with the embedded flux tube occurs. This value of $q_{cr}$ depends on the coupling constant $\alpha$.
In the limit $\alpha \to 0$ the function $\rho_2$ converges uniformly to the null function and the
branch of twisted solutions bifurcates into the local string solutions.

Here, we consider the extended non-Abelian Higgs model with our potential (\ref{pot1}) 
and find that analogous string solutions exists when the potential parameters, $\alpha,\beta$, 
are chosen large enough. A similar critical phenomenon occurs, and in addition to the critical twist 
there exists now a critical $\delta$. 

First, we study the case of an $SU(2)$ gauge group and a global $SU(3)$. We find
twisted semilocal strings corresponding to different choices of the 
 magnetic parameters $(m_1,m_2)$.  Fig. \ref{figsStaticTWSU2} shows the 
 field profiles of two typical static twisted semilocal strings with different windings and therefore 
different fluxes. Note the additional scalar field $\rho_2$ which decreases slowly, yet still 
exponentially. The relations among the currents, Eq. (\ref{SumChargesAndCurrents})
yield in this case $I_1=0$ (see explanation below (\ref{SumChargesAndCurrents})) and   
we observe that it is consistent with the numerical results 
giving $A_z =  -\tilde A_z$ for this case  as is indeed seen in Fig. \ref{figsStaticTWSU2}. 

\begin{figure}[!t]
\centering
\leavevmode\epsfxsize=10.0cm
\epsfbox{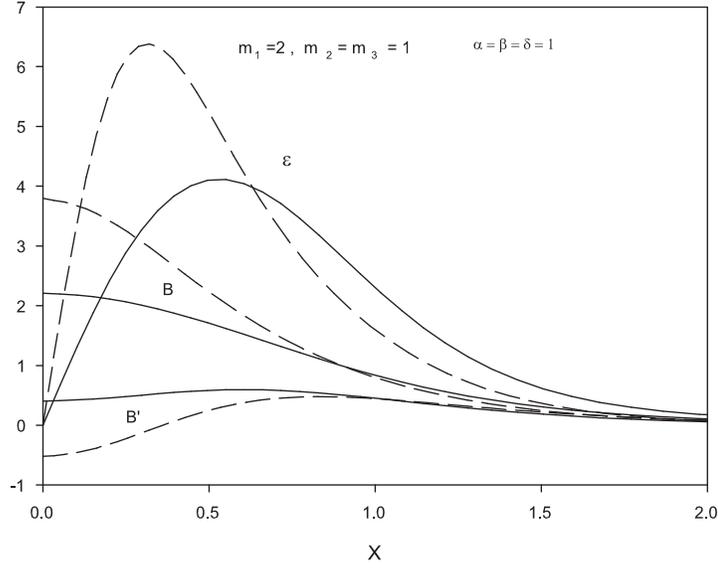}\\
\caption{\label{fig12} \small{The energy density and magnetic fields along the string axis ($B'$ stands for $\tilde{B}$)
are superposed for a twisted semilocal string $\rho_{3}(0)=1.25$ or $q= 0.34$ (solid lines) and an embedded local 
one (dashed lines).}}
\end{figure}

Analogous twisted solutions were obtained for $N=3$ and $N_f=4$.  Fig. \ref{fig6} depicts a typical solution 
corresponding to the specific Cartan generator fixed by $\psi = \pi/6$.  
The magnetic fields along the string axis (Abelian and non Abelian) and the energy density of a
twisted solution (with $\rho_{3}(0)=1.25$ or $q= 0.34$) are compared in Fig. \ref{fig12}
with those of an embedded local string. Note that the twisted strings have a non-vanishing tangential
magnetic field ($B_\theta$--see figure \ref{MagneticField}) whose contribution to the flux vanishes.
 
\begin{figure} [!t]
\centering
\leavevmode\epsfxsize=10.0cm
     \includegraphics[height=.25\textheight,width=.48\textwidth]{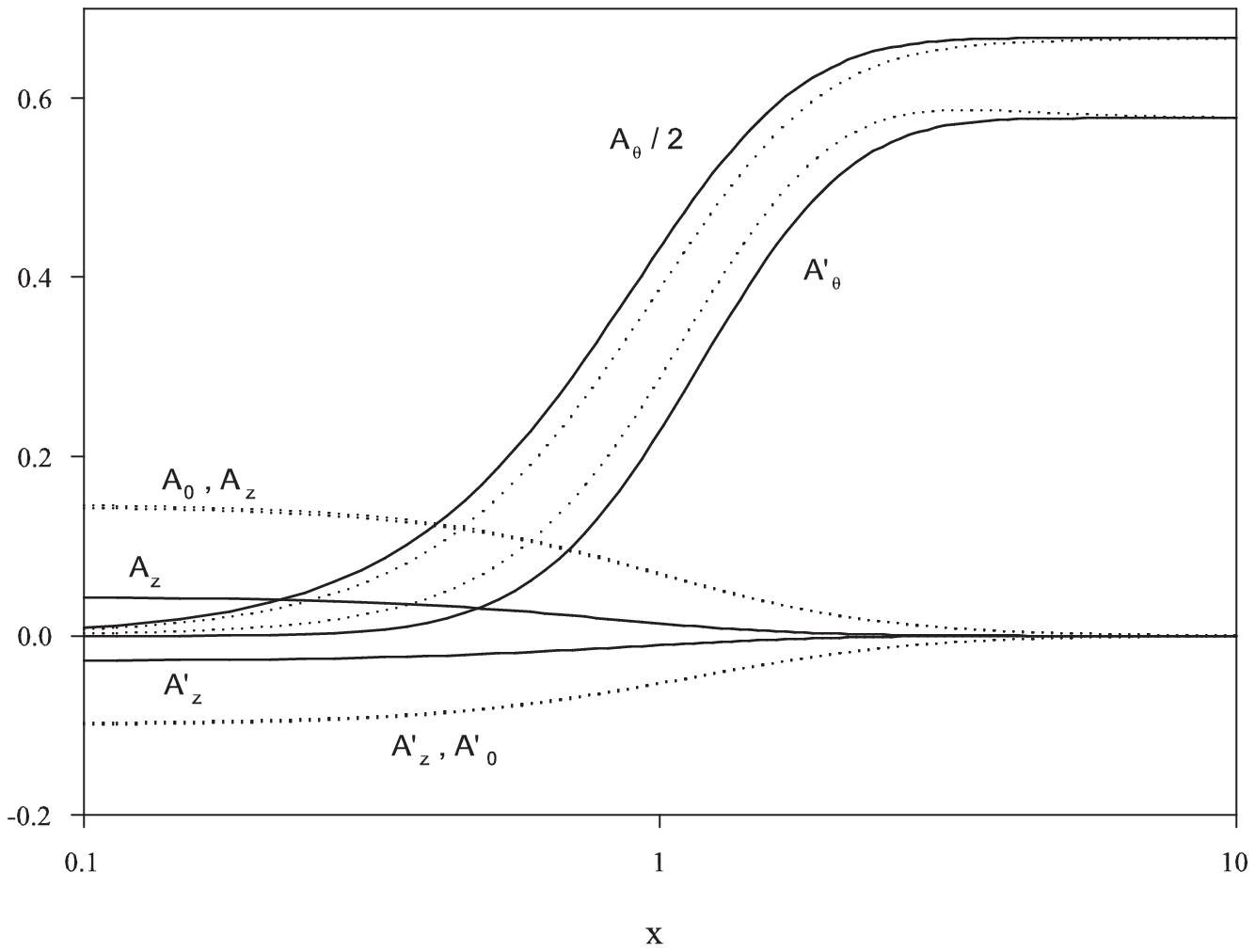}  
      \includegraphics[height=.25\textheight,width=.48\textwidth]{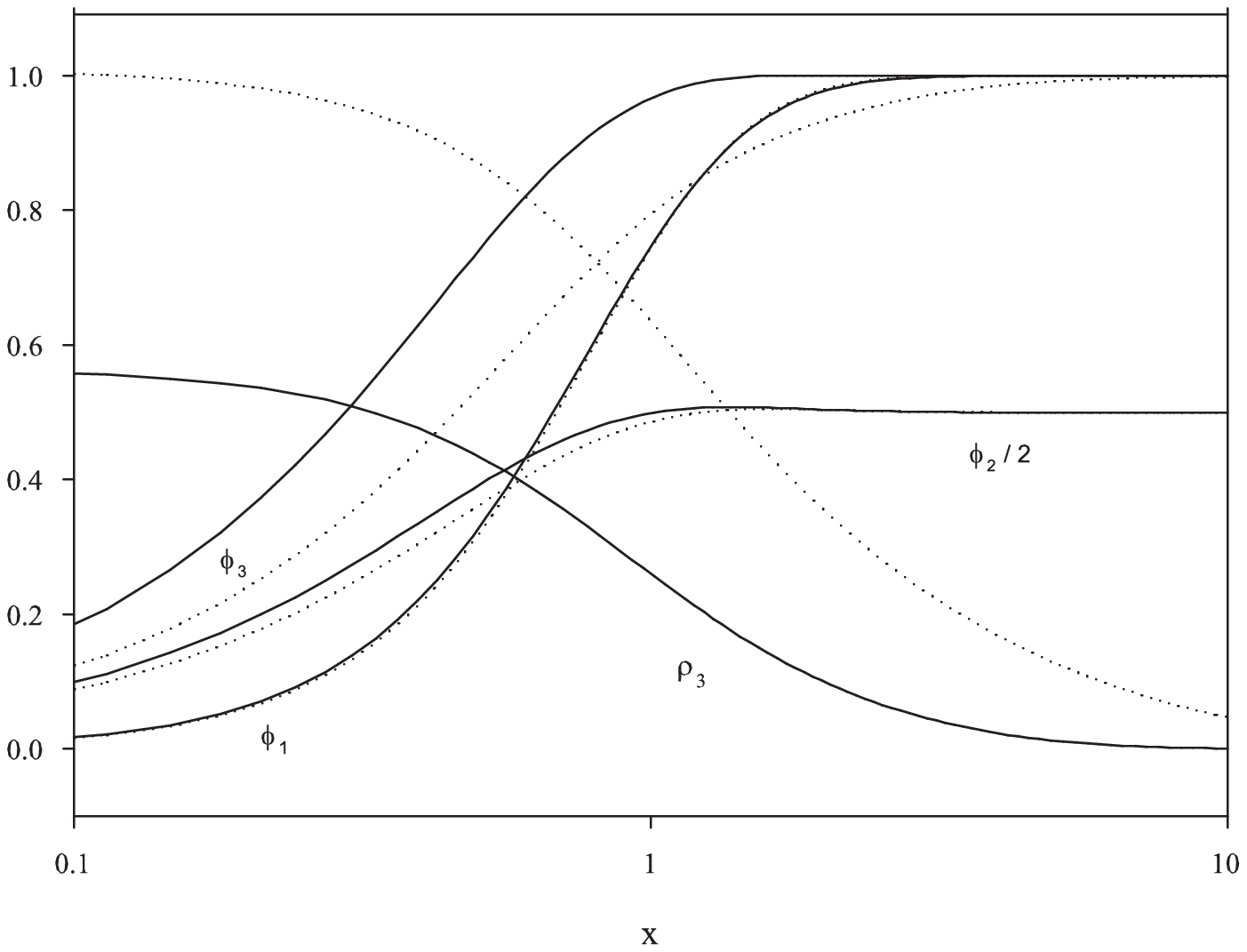}\\
\caption{\label{fig7} 
\small{Field profiles ($A'$ stands for $\tilde{A}$) of two $N = 3$ twisted solutions 
with $\alpha = \beta = \delta =1$ , $\vec m = (2,1,1)$, $q=0.5$.
Solid lines: static ($\varpi=0$); dashed lines: stationary with $\varpi=0.495$.}}
\end{figure}

The  solutions constructed above are static but they can be made
stationary by applying a boost along the string axis ($z$). Alternatively, stationary solutions 
can be obtained directly by setting the parameter $\varpi$ non zero in the field equations. 
Interestingly, allowing the parameter $\varpi\neq 0$ confers the strings with angular momentum
as was first demonstrated in the Abelian case \cite{FRV1,FRV2}. The resulting angular momentum
is therefore more like the ``orbital'' kind rather than an intrinsic spin-like. 

The profiles of  twisted solutions corresponding to $m_1=2,m_2=m_3=1$
and   $q=0.5$ are presented in Fig. \ref{fig7} for  $\varpi=0$ and $\varpi=0.495$ (i.e.
very close to the chiral limit $\varpi=q$). 
We note that the chiral limit $\varpi\rightarrow q$ leads to a 
different kind of solutions with $\rho_3$ decaying as a 
power instead of exponentially. This tendency is apparent 
in Fig. \ref{fig7} namely, through the decay of the function $\rho_3$.

\begin{figure}[!b]
\centering
\leavevmode\epsfxsize=10.0cm
\epsfbox{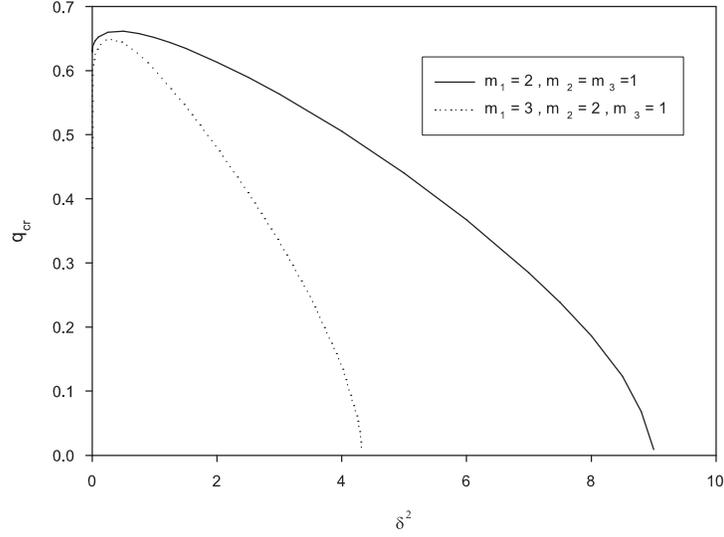}\\
\caption{\label{fig9} \small{The critical $q$-value as a 
function of the coupling ratio $\delta^2$ for $\alpha=\beta=1$.}}
\end{figure}

\subsection{Analysis of parameter space}

Since we know from the Abelian case ($N=1$) that twisted semilocal strings exist only
is a restricted domain  of the parameter space, it is natural to try to 
figure out this domain in the present non-Abelian model. The Lagrangian (\ref{lag}) is characterized by
four parameters: $\alpha$, $\beta$ of the scalar potential and two gauge coupling constants
$e_1$ and $e_2 $. However, only the relative strength $\delta =e_2 /e_1$ appears in the field
equations so the parameter space of the model is three-dimensional. 

\begin{figure}[!b]
\centering
\leavevmode\epsfxsize=10.0cm
\epsfbox{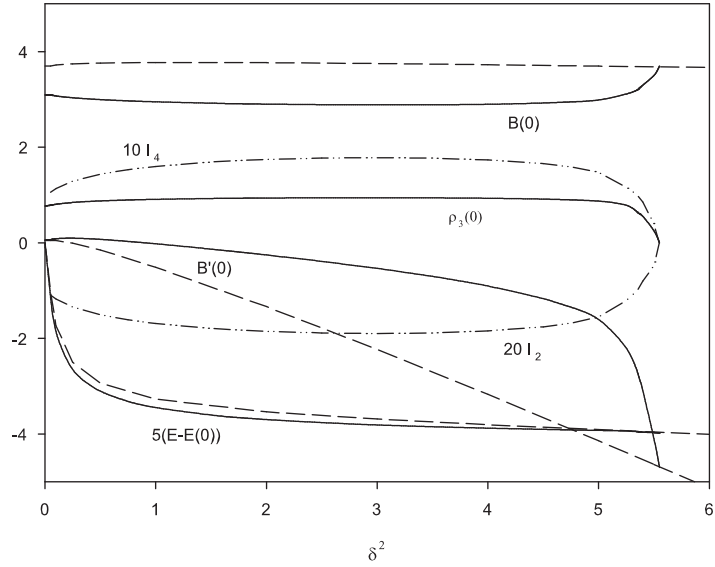}\\
\caption{\label{fig10} \small{Energy, global currents (see (\ref{GlobalCurrents})), 
central magnetic fields ($B'$ stands for $\tilde{B}$) and
central ``twisted'' scalar field as a function of the coupling ratio $\delta^2$ for static twisted 
semilocal strings with $q=0.4$ (solid lines)
 and embedded local ones (dashed lines).  Both for 
 $\alpha = \beta =1$ and $\vec m = (2,1,1)$.
Note that the twisted string has lower energy.}}
\end{figure}
 
\begin{figure}[!b]
\centering
\leavevmode\epsfxsize=10.0cm
\epsfbox{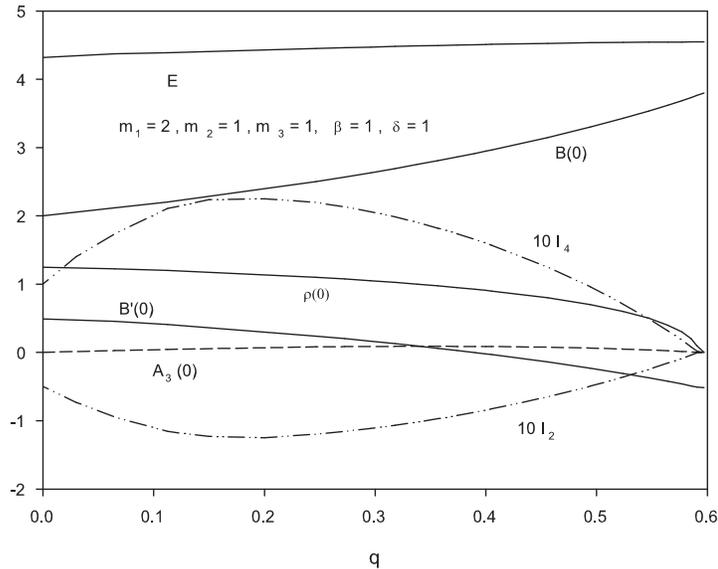}\\
\caption{\label{fig11} \small{Energy, global currents, central magnetic fields ($B'$ stands for $\tilde{B}$),
 gauge potential and scalar field as function of the twist parameter for $\alpha = \beta = \delta =1$ for $\vec m = (2,1,1)$. 
Note that $q=0$ is not included.}}
\end{figure}

The parameter space of the {\it solutions} is far more complicated since it contains also the twist $q$, 
the other phase parameter $\varpi$ and the flux numbers.

We investigated this problem in the case of an $SU(3)$ gauge group and we present here the main results in
various planes in this higher dimensional parameter space. We took representative
values of $\alpha=\beta=1$ for which it turns out that twisted semilocal strings exist.

The effect of  changing the parameter $\delta$ on the twisted solutions is illustrated in Fig. 
\ref{fig9} where the critical $q_{cr}$-value is plotted as a function of $\delta^2$ 
for two different choices of $\vec m = (m_1 , m_2 , m_3)$. The choice $(2,1,1)$ corresponds to
$\psi = \pi/6$ and the choice $(3,2,1)$ to $\psi = \pi/3$ as explained at the end of Sec. \ref{secFEqs}. 
The twisted semilocal strings exist only for $q < q_{cr}$ and for values of $\delta^2$ 
below a maximal value of $\delta_{cr}^2$ . This value depends on $\vec m$ and
determines a domain of existence of twisted solutions in the $(\delta,q)$ plane.

In Fig. \ref{fig10}, a few physical quantities characterizing the solutions corresponding to $q=0.4$
are presented. We find $\delta_{cr}^2 =5.55$ which is consistent with the value that 
is obtained from Fig. \ref{fig9}. We further studied the critical phenomenon 
appearing when the  maximal value of $\delta$ (with $q$ fixed) is reached as seen in Fig. \ref{fig10}.   
 Our numerical results suggest in particular  that the twisted solution bifurcates into 
 the embedded local string  at $\delta = \delta_{cr}$. Note especially the vanishing of $I_2$, $I_4$ 
 and $\rho_3 (0)$ as $\delta \to \delta_{cr}$.

The evolution of the physical parameters characterizing the twisted solutions for $\delta=1$
is further presented in Fig. \ref{fig11} as a function of the twist parameter $q$. 
We see from the plot that the solution bifurcates into the embedded local string
for $q = 0.597$ which is of course consistent with $q_{cr}(1)$ of Fig. \ref{fig9}. 
We notice that  the central value of the non Abelian magnetic field, $\tilde{B}(0)$ 
changes sign between the bifurcation point and the low twist region (i.e. $q \to 0$).

\begin{figure}[!t]
\centering
\leavevmode\epsfxsize=10.0cm
\epsfbox{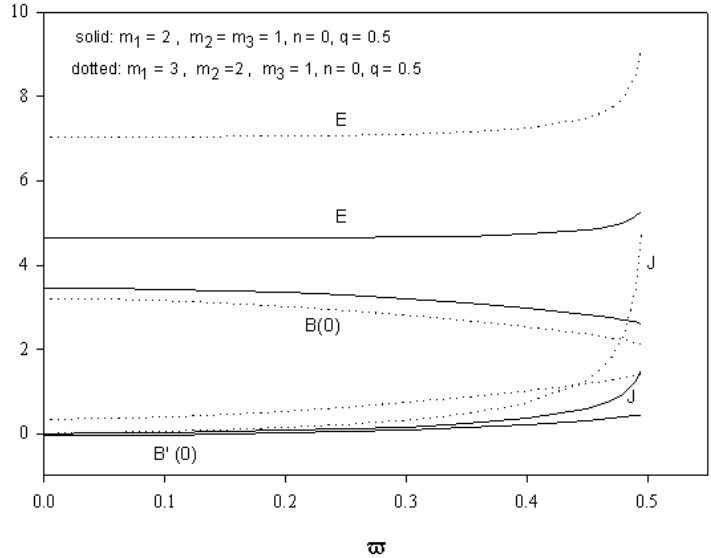}\\
\caption{\label{fig8} \small{Evolution of the Mass and the 
magnetic fields ($B'$ stands for $\tilde{B}$) on the axis as functions of $\varpi$}}
\end{figure}

The  effect of a boost along the string axis, which is reflected by a non-vanishing $\varpi$,  
on some physical quantities is presented in Fig. \ref{fig8}. In this figure
$E,J,B(0)$ and $\tilde{B}(0)$ are plotted versus $\varpi$ for 
a fixed value of $q$ (for instance $q=0.5$) and for the two values  
$\psi = \pi/6\;,  \pi/3$ which require different sets of $m_a$ as explained above. 
We stress that the plots are made for fixed $q$, so the
solutions involved are not related to each other
by boosts along the $z$ direction. In particular, the magnetic fields
on the $z$-axis vary with $\varpi$. Note also that the chiral limit has nontrivial 
consequences on the mass and spin of the solutions (per unit length) which diverge in this limit
 as indicated  in Fig. \ref{fig8}.
    
\section{Conclusions}
\setcounter{equation}{0}

This work presents string solutions in the extended non-Abelian Higgs model which may be 
seen as an expansion of recent discussions of Abelian twisted semilocal strings \cite{FRV1,FRV2}
and of non-Abelian untwisted ones \cite{IsozumiEtAlPRD2005,ShifmanYung2006,GorskyMikh2007,AuzziEtAl2007}. 
The solutions are classified according to the
twist parameter $q$ as twisted or untwisted (for $q=0$). The twisted solutions are characterized by
a persistent current along the string axis. The twisted solutions are further classified
as static ($\varpi=0$) or stationary, where the latter are also characterized by a non-zero 
angular momentum. The extended non-Abelian Higgs model allows also two kinds of special solutions 
-- both untwisted and both purely magnetic: The embedded local non-Abelian flux tubes and the self-dual
 ``Skyrmions''. 
 
The occurence of Twisted semilocal string and their corresponding conserved currents is closely connected 
to the existence of a scalar field which condensates on the axis and vanishes asymptotically. 
For $N_f=N$ all scalars are Higgs-like fields which acquire asymptotically a non vanishing expectation value due to the
 potential. As a consequence, there is no place for a condensate  and no twisted semilocal string are possible 
 in this case. 
 
All those solutions were constructed by numerically solving the field equations with
the appropriate boundary conditions.  The solutions exist only
on a non trivial domain of the parameter space of the possible solutions.
We have tried to determine this domain qualitatively and 
obtained a lot of information about the pattern of the solutions. We demonstrated the bifurcation
phenomenon which occurs in this case at a critical value of the twist $q$ which now depends on the various
fluxes and on $\delta = {e_2}/{e_1}$ in addition to the parameters in the potential.

A crucial step for deciding the physical relevance of non-Abelian twisted semilocal strings would be to 
study their stability by performing a systematic normal mode analysis, as was done recently \cite{garaud+volkov} 
for the extended Abelian Higgs model. Another important issue is the understanding of the low-energy
 dynamics of the non-Abelian twisted semilocal strings. This can be done e.g. by using the geodesic approximation
 of \cite{Manton1982,TopolSol} which is based on a parametrization of the zero modes in
 terms of collective coordinates. For untwisted local and semilocal strings, this
 analysis is reported in details by Aldrovandi (see sec. VI of \cite{Aldrovandi2007}). 
 Since the phases of twisted semilocal strings depend on $t$ and $z$, the parametrization
 of the orientational moduli in terms of slowly varying functions  of these coordinates does 
 not seem to be straightforward.  The parametrization of the fluctuations used in \cite{garaud+volkov}
  might turn out to be very useful. We plan to address these questions in a further publication, as well
 as the gravitating counterparts of twisted strings.  

Another direction is motivated by the non-Abelian superconducting strings, recently obtained by Volkov
\cite{Volkov2006} within the $SU(2) \times U(1)$ electroweak theory. These ``untwisted'' strings 
can be used as a starting point towards constructing twisted solutions which will be related to a 
non-Abelian subgroup of the gauge group.

\noindent{\bf Acknowledgments:}
We gratefully acknowledge Prof. M. Nitta for useful comments and information about some 
relevant papers. One of us (Y. V.) gratefully acknowledges the Belgian F.N.R.S. for a grant supporting 
a visit at the University of Mons-Hainaut where most of this work was carried out.


\begin{thebibliography}{8}
\bibitem{NO} H. B. Nielsen and P. Olesen, Nucl. Phys. {\bf B 61} (1973) 45. \vspace{-0.2cm}
\bibitem{deVega1976} H.~J.~de Vega,
  Phys.\ Rev.\  D {\bf 18}, 2932 (1978). \vspace{-0.2cm}
\bibitem{Hasenfratz1979}
  P.~Hasenfratz,
  Phys.\ Lett.\  B {\bf 85}, 338 (1979). \vspace{-0.2cm}
\bibitem{SchwarzTyupkin}
  A.~S.~Schwarz and Yu.~S.~Tyupkin,
  Phys.\ Lett.\  B {\bf 90} (1980) 135. \vspace{-0.2cm}
\bibitem{deVegaSchap1986PRL}
  H.~J.~de Vega and F.~A.~Schaposnik,
  Phys.\ Rev.\ Lett.\  {\bf 56}, 2564 (1986). \vspace{-0.2cm}
  \bibitem{deVegaSchap1986PRD}
  H.~J.~de Vega and F.~A.~Schaposnik,
  Phys.\ Rev.\  D {\bf 34}, 3206 (1986). \vspace{-0.2cm}
\bibitem{HeoVach1998}
  J.~Heo and T.~Vachaspati,
  Phys.\ Rev.\  D {\bf 58}, 065011 (1998). \vspace{-0.2cm}
\bibitem{Suranyi1999}
  P.~Suranyi,
  Phys.\ Lett.\  B {\bf 481}, 136 (2000). \vspace{-0.2cm}
\bibitem{SchaposnikSuranyi}
  F.~A.~Schaposnik and P.~Suranyi,
  Phys.\ Rev.\  D {\bf 62}, 125002 (2000). \vspace{-0.2cm}
  \bibitem{KneippBrockill}
  M.~A.~C.~Kneipp and P.~Brockill,
  Phys.\ Rev.\  D {\bf 64}, 125012 (2001). \vspace{-0.2cm}
  \bibitem{KonishiSpanu}
  K.~Konishi and L.~Spanu,
  Int.\ J.\ Mod.\ Phys.\  A {\bf 18}, 249 (2003). \vspace{-0.2cm}
\bibitem{VilShel} A. Vilenkin and E.P.S. Shellard, \textit{Cosmic Strings and Other
Topological Defects}, (Cambridge University Press, Cambridge,
England, 1994). \vspace{-0.2cm}
\bibitem{HananyTong2003}
  A.~Hanany and D.~Tong,
  JHEP {\bf 0307}, 037 (2003). \vspace{-0.2cm}
\bibitem{Auzzi2003}
  R.~Auzzi, S.~Bolognesi, J.~Evslin, K.~Konishi and A.~Yung,
  Nucl.\ Phys.\ B {\bf 673}, 187 (2003). \vspace{-0.2cm}
\bibitem{ShifmanYung2004}
  M.~Shifman and A.~Yung,
  Phys.\ Rev.\  D {\bf 70}, 045004 (2004). \vspace{-0.2cm}
\bibitem{HananyTong2004}
  A.~Hanany and D.~Tong,
  JHEP {\bf 0404}, 066 (2004). \vspace{-0.2cm}
  \bibitem{MarkovEtAl}
  V.~Markov, A.~Marshakov and A.~Yung,
  Nucl.\ Phys.\  B {\bf 709}, 267 (2005). \vspace{-0.2cm}
\bibitem{EtoEtAlPRL2006}  
M.~Eto, Y.~Isozumi, M.~Nitta, K.~Ohashi and N.~Sakai,
  Phys.\ Rev.\ Lett.\  {\bf 96}, 161601 (2006). \vspace{-0.2cm}
\bibitem{EtoEtAlJPA2006}
  M.~Eto, Y.~Isozumi, M.~Nitta, K.~Ohashi and N.~Sakai,
  J.\ Phys.\ A  {\bf 39}, R315 (2006). \vspace{-0.2cm}
\bibitem{EtoEtAlPRD2006}
  M.~Eto, K.~Konishi, G.~Marmorini, M.~Nitta, K.~Ohashi, W.~Vinci and N.~Yokoi,
  Phys.\ Rev.\  D {\bf 74}, 065021 (2006). \vspace{-0.2cm}
\bibitem{IsozumiEtAlPRD2005}
  Y.~Isozumi, M.~Nitta, K.~Ohashi and N.~Sakai,
  Phys.\ Rev.\  D {\bf 71}, 065018 (2005). \vspace{-0.2cm}
\bibitem{ShifmanYung2006}
  M.~Shifman and A.~Yung,
  Phys.\ Rev.\  D {\bf 73}, 125012 (2006). \vspace{-0.2cm}
  
\bibitem{VachAch} T. Vachaspati and A. Achucarro, Phys. Rev. {\bf  D 44} (1991) 3067. \vspace{-0.2cm}
\bibitem{AchVach} A. Achucarro and T. Vachaspati, Phys. Rep. {\bf 327} (2000) 427. \vspace{-0.2cm} 

\bibitem{GorskyMikh2007}
  A.~Gorsky and V.~Mikhailov,
  Phys.\ Rev.\  D {\bf 76}, 105008 (2007). \vspace{-0.2cm}
\bibitem{AuzziEtAl2007}
  R.~Auzzi, M.~Eto and W.~Vinci,
  arXiv:0711.0116 [hep-th]. \vspace{-0.2cm} 
 
\bibitem{FRV1} P. Forgacs, S. Reuillon and M.S. Volkov, Phys. Rev. Lett. {\bf 96} (2006) 041601. 
\vspace{-0.2cm}
\bibitem{FRV2} P. Forgacs, S. Reuillon and M.S. Volkov, Nucl. Phys. {\bf B 751} (2006) 390. \vspace{-0.2cm}
\bibitem{Volkov2006}
  M.~S.~Volkov,
  Phys.\ Lett.\  B {\bf 644}, 203 (2007). \vspace{-0.2cm}
\bibitem{Aldrovandi2007}
L.~G.~Aldrovandi,
Phys.\ Rev.\  D {\bf 76}, 085015 (2007). \vspace{-0.2cm}  
\bibitem{HashimotoTong}
  K.~Hashimoto and D.~Tong,
  JCAP {\bf 0509}, 004 (2005). \vspace{-0.2cm}
\bibitem{EtoEtAl2007}
  M.~Eto, K.~Hashimoto, G.~Marmorini, M.~Nitta, K.~Ohashi and W.~Vinci,
  Phys.\ Rev.\ Lett.\  {\bf 98}, 091602 (2007). \vspace{-0.2cm}
\bibitem{Hindmarsh1992}
  M.~Hindmarsh,
  Phys.\ Rev.\ Lett.\  {\bf 68}, 1263 (1992). \vspace{-0.2cm}
\bibitem{Preskill1992}
  J.~Preskill,
  Phys.\ Rev.\  D {\bf 46}, 4218 (1992). \vspace{-0.2cm}
   \bibitem{COLSYS}
 U. Ascher, J. Christiansen, R.~D. Russell,
 Mathematics of Computation {\bf 33} (1979) 659;
 ACM Transactions {\bf 7} (1981) 209. \vspace{-0.2cm}
\bibitem{garaud+volkov} J. Garaud and M.S. Volkov 
arXiv:0712.3589, To appear in Nucl. Phys. \vspace{-0.2cm}
\bibitem{Manton1982} N. Manton,  Phys.\ Lett.\  B {\bf 110}, 54 (1982). \vspace{-0.2cm}
\bibitem{TopolSol} N. Manton and P. Sutcliffe, \textit{Topological Solitons}, 
(Cambridge University Press, Cambridge, England, 2004). \vspace{-0.2cm} 

\end{thebibliography}
\end{document}